\documentclass[aps,prl,twocolumn]{revtex4}
\usepackage{amsmath,graphicx}

\begin{document}

\title{Scaling and the center of band anomaly
in a one-dimensional Anderson model with diagonal disorder}
\author{L. I. Deych$^a$}
\author{M. V. Erementchouk$^a$}
\author{A. A. Lisyansky$^a$}
\author{B. L. Altshuler$^b$}
\address{$^a$Physics Department, Queens College of the City University of New York, Flushing, NY 11367\\
$^b$Physics Department, Princeton University and NEC Laboratories
America, Princeton, NJ 08540}

\begin{abstract}
We resolve the problem of the violation of single parameter
scaling at the zero energy of the Anderson tight-binding model
with diagonal disorder. It follows from the symmetry properties of
the tight-binding Hamiltonian that this spectral point is in fact
a boundary between two adjacent bands. The states in the vicinity
of this energy behave similarly to states at other band
boundaries, which are known to violate single parameter scaling.
\end{abstract}

\maketitle

\section{Introduction}

Despite a long history of research in the field of localization,
even such a standard problem as the one-dimensional Anderson
tight-binding model with nearest neighbor hopping and diagonal
disorder can still hold surprises. This model is described by the
equation of motion
\begin{equation}
\psi _{n+1}+\psi _{n-1}+(\epsilon _{n}-E)\psi _{n}=0,  \label{eq_of_motion}
\end{equation}
where, $\psi _{n}$ represents a wave function at the $n$-th site,
$\epsilon _{n}$ is a random site energy distributed uniformly
between $-W$ and $W$, and $E$ is the energy. Recently, Schomerus
and Titov \cite{Titov} demonstrated that single parameter scaling
(SPS), which is a cornerstone of the conceptual framework for
dealing with transport properties of disordered systems
\cite{gang}, is violated in this classical model of localization
in the vicinity of the energy $E=0$. While the violation of SPS in
the vicinities of the energies $E=\pm 2$ (which are band
boundaries of the system with $W=0$, i.e., all $\epsilon _{n}=0$)
is well established \cite{DeychPRL}, the absence of SPS at the
band center $E=0$ looks surprising. It is yet another
manifestation of the anomalous nature of this spectral point.
Whereas earlier studies of the Anderson model revealed an
anomalous behavior of the localization length at $E=0$
\cite{Kappus,Derrida,Goldhirsch1994,Izrailev}, the violation of
SPS in this spectral region remained unnoticed. The mathematical
roots of the anomalous behavior of the localization length were
established in Ref.~\onlinecite{Goldhirsch1994}. At the same time
the physical understanding of the nature of the anomalous
properties of $E=0$ is still absent. Here we suggest an approach
to this spectral region based on symmetries of the Hamiltonian
Eq.~(\ref{eq_of_motion}), which clarifies the physical nature of
this anomaly, and explains scaling properties of the Anderson
model in this spectral region.

It is convenient to describe the statistics of conductance, $g$,
of a disordered chain of length $L$ in terms of the Lyapunov
exponent (LE), $\tilde{\gamma}$, defined as \cite{Anderson}
\begin{equation}
\tilde{\gamma}=\frac{1}{L}\ln \left( 1+\frac{1}{g}\right).
\label{LE}
\end{equation}
SPS means that the probability distribution of $\tilde{\gamma}$
depends on a single parameter, so that, e.g., any moment of this
distribution can be expressed through the mean value of LE,
$\gamma \equiv \langle\tilde{\gamma}\rangle$. Such a relationship
between $\gamma$ and the variance, $\sigma ^{2} \equiv
\langle\tilde{\gamma}^2\rangle-\langle\tilde{\gamma}\rangle^2$,
was first conjectured in Ref.~\onlinecite{Anderson}
\begin{equation}
\tau \equiv \sigma^{2}L/\gamma =1.  \label{SPS}
\end{equation}%
The authors of Ref.~\onlinecite{Titov} demonstrated that at $E=0$,
parameter $\tau $ of Eq.~(\ref{SPS}) deviates from the SPS value
and is equal to $1.047$. This deviation from  Eq.~(\ref{SPS}),
which was overlooked by previous studies probably due to its small
size, is of principle importance. It seemingly contradicts  a well
established criterion for SPS introduced in
Ref.~\onlinecite{DeychPRL}.

In this Letter we show that this is only an apparent
contradiction, and explain the violation of SPS at $E=0$ within
the general picture developed in Ref.~\onlinecite{DeychPRL}. The
anomaly at $E=0$ turns out to result from a hidden symmetry of the
tight-binding model, which splits the conduction band into two
adjacent bands with a boundary at $E=0$. This splitting does not
affect thermodynamic properties of the model but is crucial to its
scaling behavior.

Before presenting our resolution of the mysterious behavior in the
vicinity of $E=0$, let us recall the history of the criterion for
SPS. Originally, Anderson, et al.~\cite{Anderson} suggested that
SPS holds when the localization length of the system,
$l_{loc}=\gamma ^{-1},$ exceeds what they called the phase
randomization length: $l_{loc}>l_{ph}$. The stationary
distribution of the phases of the reflection and transmission
amplitudes in disordered one-dimensional systems was assumed to
become uniform as soon as the system length $L$ exceeds $l_{ph}$.
Many authors derived Eq.~(\ref{SPS}) from this phase randomization
hypothesis for several different models. The phase randomization
itself was rigorously proven only for few particular models:
banded matrices \cite{Mirlin}, the Anderson tight-binding model
with diagonal disorder \cite{Goldhirsch1994}, and a continuous
model with a white-noise random potential \cite{Schomerus} for
certain spectral regions. At the same time, it was shown that for
some values of energies, such as $E=0$ or $E=\pm 2$ for the model
described by Eq.~(\ref{eq_of_motion})
the stationary distribution of the phases is not uniform \cite%
{Stone,Kappus,Derrida,Goldhirsch1994,Izrailev,Ossipov}.

Although in the vicinities of these special energies SPS is indeed
violated \cite{ZaslavskyPRL,Titov}, the criterion for SPS based on
the phase randomization length does not seem to describe the
situation adequately. Indeed, the absence of the phase
randomization in all situations discussed above does not mean that
the \textquotedblleft relaxation\textquotedblright\ length of the
phase distribution diverges and exceeds the localization length
when one approaches the respective energies. What actually happens
is that the uniform contribution to the stationary phase
distribution vanishes in the vicinity of these special energies.
As a result, the nonuniform part, which always exists but is
usually small, becomes dominant
\cite{Derrida,Goldhirsch1994,Izrailev}. Thus, the phase
randomization length, as it was introduced in
Ref.~\cite{Anderson}, is ill-defined.

Thus, appealing to the phase distribution in connection with SPS
simply substitutes the problem of formulating a criterion for SPS
by the problem of finding a criterion for the uniform distribution
of phases. Previously proposed criteria for the latter are
specific for particular models. For instance, in
Ref.~\onlinecite{Titov} for the Anderson model within the white
noise approximation, the criteria for the phase randomization for
the band edges and the band center are set as $|E-2|>3W^{2/3}$ and
$|E|>10D$, respectively ($D$ is a measure of the disorder
strength). These definitions do not look satisfactory to us. We
believe that the criterion should be universal and formulated in
terms of some fundamental macroscopic quantities.

Such a universal criterion for SPS similar in form to the
Anderson's, $l_{loc}>l_{ph}$, was recently suggested in
Ref.~\onlinecite{DeychPRL}:
\begin{equation}
\kappa \equiv l_{loc}/l_s > 1.  \label{eles}
\end{equation}
The new length scale, $l_s$, has nothing to do with $l_{ph}$ and
can be expressed through the integral density of states normalized
by the total number of states in the band, $N(E)$, and a distance
between the neighboring sites, $a$:
\begin{equation}
l_{s}=a/\sin[\pi N(E)].  \label{ls}
\end{equation}
This criterion was extracted from the exact calculation of the
variance of LE for the Anderson model with the Cauchy distribution
of the site energies, $\epsilon _{n}$ (the Lloyd model).
Eq.~(\ref{SPS}) was derived in Ref.~\onlinecite{DeychPRL} without
any relation to the phase distribution \footnote{The actual result
for $\tau $ obtained in Ref.~\onlinecite{DeychPRL} differs by the
factor of $2$ from the standard form Eq.~(\ref{SPS}) because of
the peculiarities of the Cauchy distribution}.  For in-band
states, $N(E)\sim 1/2$ and $l_s$ is microscopic and therefore is
not significant. For the states close to the band edges and
fluctuation states in the former band gaps, where $N(E)\ll 1$ or
$1-N(E)\ll 1$, $l_s$ becomes macroscopic and can exceed $l_{loc}$.
In the latter case, $l_s$ has a clear physical meaning as an
average distance between localization centers responsible for the
states with energies between $E$ and the closest fluctuation
boundary of the spectrum.

The expression of $l_{s}$ in terms of the density of states $N(E)$
allowed authors of Refs.~\onlinecite{DeychPRL} to conjecture a
generalization of this definition to other models. It was
suggested that $N(E)$ in Eq.~(\ref{ls}) must be understood as the
total number of states (per unit length) between $E$ and a closest
genuine spectral boundary normalized by the total number of
states. This definition of $l_{s}$ can be used also in the case of
systems with multiple bands, provided that the disorder does not
fill the band gaps of the original systems completely, and there
exist genuine spectral boundaries inside each of the band gaps.
Under this condition, $N(E)$ refers to the number of states
associated with one particular band of the spectrum, and varies
between $0$ and $1$ when $E$ spans the states of the band.

In light of this criterion, one can understand the violation of
SPS in the band gaps of the initial spectrum observed numerically
in Ref.~\onlinecite{ZaslavskyPRL}. Applicability of the
$l_{s}$-based criterion was examined for several other models such
as the Anderson model with the box \cite{DeychPRL} and dichotomic
\cite{Erementchouk} distributions of the site energies, the model
of a scalar wave propagating through a one-dimensional disordered
periodic-on-average superlattice, and even for the model of a wave
propagating in an absorbing medium \cite{YamilovPRB}. In all these
cases, the criterion perfectly described the transition between
the SPS behavior of in-band states, and non-SPS behavior of the
band edge and fluctuation states. Moreover, as  was shown recently
in Ref.~\onlinecite{DeychPRL2002}, the length $l_{s}$ plays an
even more important role than simply determining a boundary
between SPS and non-SPS spectral regions: the second and the third
moments of the distribution of the Lyapunov exponent in the
non-SPS region can be parameterized by a single parameter $\kappa$
provided that $L$ exceeds $l_{s}$ \footnote{For intermediate
lengths, $l_{loc}\ll L\ll l_{s}$, there is an additional parameter
$L/l_{s}$, which describes an anomalous behavior of the second
moment \cite{DeychPRL2002}.}.

\section{The band-center anomaly in the Anderson model}

The success of the scaling approach based on the parameter
$\kappa$, Eqs.~(\ref{eles}) and (\ref{ls}), in predicting
violations of SPS at the band edges of different models, and in
describing the probability distribution of LE in non-SPS spectral
regions \cite{DeychPRL,YamilovPRB,DeychPRL2002} motivated us to
look more carefully at the reasons for the apparent failure of
this approach at the center of the band of the Anderson model.

We start our analysis with a tight-binding model without disorder,
$\epsilon_{n}=0$. The solution of  Eq.~(\ref{eq_of_motion}) at
$\epsilon_{n}=0$ is plane waves, $\psi _{n}\propto \exp (\pm
ikna)$, with  $k$ satisfying the dispersion equation
\begin{equation}
E=2\cos {ka};\hspace{0.3in}0\leq ka\leq \pi ,  \label{dispersion}
\end{equation}
According to the traditional widely accepted point of view,
Eq.~(\ref{dispersion}) describes a single energy band $-2\leq
E\leq 2$,

However, this model contains more than the simple picture reveals.
Indeed, the equations of motion, Eq.~(\ref{eq_of_motion}), at
$\epsilon_{n}=0$ acquire an additional symmetry: Operation
\begin{equation}
D_{n}=\psi _{n}\rightarrow (-1)^{n}\psi _{n} \label{Dn}
\end{equation}
transforms a state $\psi _{n}(E)$ into the state $\psi _{n}(-E)$:
$\psi _{n}(E)=(-1)^{n}\psi _{n}(-E)$. This operation forms a
group, and the states of this model can be classified according to
its irreducible representations.

In order to realize these representations, we introduce two field
invariants with respect to $D_{n}$:
\begin{equation}
u_{l}=\psi _{2l};\hspace{0.3in}v_{l}=\psi _{2l+1}.  \label{representations}
\end{equation}
Solving equations of motion for these fields we find that the
spectrum is separated into two branches described by the following
dispersion equations
\begin{equation}
E=\pm \cos (ka/2). \label{newdisp}
\end{equation}
Two bands of the spectrum described by Eq.~(\ref{newdisp}) have a
common boundary at $E=0$, and, therefore, the spectrum of the
system appears as consisting of a single band, as it is usually
assumed. However, the classification of the states according to
the irreducible representations of $D_{n}$ reveals that $E=0$
should be considered as a band boundary rather than as a band
center. The new description of the band structure can be viewed as
a transition to the reduced zone representation, where the wave
number of the original dispersion equation Eq.~(\ref{dispersion})
is restricted to the interval $0\leq ka\leq \pi /2$. This
operation is equivalent to doubling of the elementary cell of the
original periodic chain. This can be justified by noticing that
two adjacent sites of the chain are not equivalent with respect to
the operation $D_{n}$ [Eq.~(\ref{Dn})].

The considerations presented above would have been merely an
exercise, if the vicinity of $E=0$ did not behave anomalously in
the presence of disorder. Of course, any disorder mixes states
from $E<0$ and $E>0$ bands. However, since violation of SPS occurs
in the vicinity of a spectral boundary of an unperturbed system
\cite{DeychPRL}, one can expect that $E=0$ behaves similarly to
$E= \pm 2$ boundaries. Actually a similarity between properties of
$E=0$ and other band boundaries was noted already in
Ref.~\onlinecite{Goldhirsch1994}, where it was shown that the
calculation of the LE in the vicinities of all three points $E=\pm
2$, and $E=0$ required almost identical mathematical approach.

The deviation of the parameter $\tau (\kappa )$, Eq.~(\ref{SPS}),
from unity observed in Ref.~\onlinecite{Titov} can be interpreted
as a manifestation of the band-edge nature of the spectral point
$E=0$. This deviation is, however, different from the behavior of
$\tau (\kappa)$ in the vicinity of $E=\pm 2$. Near $E=\pm 2$,
$\tau (\kappa)$ demonstrates a small overshoot above unity at the
band side of these boundaries, and decreases quickly at the band
gap sides \cite{DeychPRL}. Since two conduction bands are adjacent
at $E=0$, there is no band gap related decrease of $\tau (\kappa
)$. However, the band-edge nature of $E=0$ still manifests itself
in the form of the overshoot, discovered in
Ref.~\onlinecite{Titov}. This overshoot appears at  both sides of
$E=0$.

The fundamental question is whether the deviation from SPS near
$E=0$  occurs in accordance with the criterion Eq.~(\ref{eles})?
The answer is ``yes'', however, the definition of the length
$l_s$, Eq.~(\ref{ls}), should be modified to reflect the fact that
the states are now distributed between two bands.  Accordingly,
when calculating $l_s$ for each of the bands, we have to keep in
mind that $N(E)$ in Eq.~(\ref{ls}) is a number of states between
$E=0$ and a given energy $E$ normalized by the total number of
states in this band.

\begin{figure}[tbp]
\vspace{-0.4in} \hspace{-0.35in} % Requires \usepackage{graphicx}
\includegraphics[width=3.7 in,angle=-0]{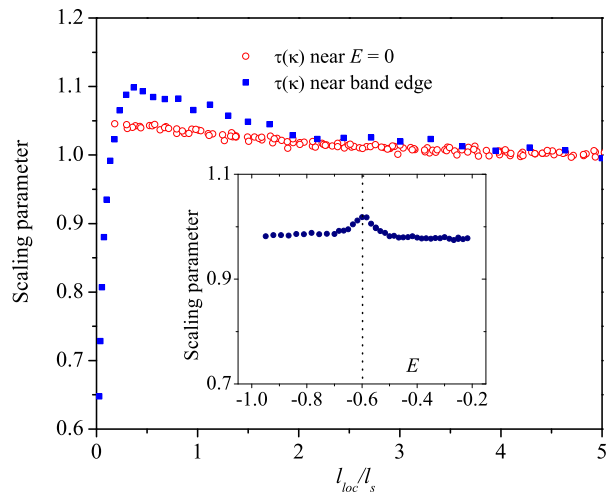}
\vspace{-0.4in} \caption{Dependence of the scaling parameter
$\tau$ on $\kappa$ in the vicinity of the fluctuating band edge
$|E|\sim 2$ and near $E=0$ for different strengths of disorder:
$W$ changes from $0.15$ to $0.4$. Curves corresponding to
different strengths are indistinguishable. The inset shows
$\tau(\kappa)$ for the Anderson model with next to the nearest
neighbors interaction, Eq.~(\ref{eq_of_motion1}), in the vicinity
of the energy $E = -2\alpha$ for $\alpha=0.3$.}
\label{fig:E0scaling}
\end{figure}

 The results of the numerical calculations based on the new
definition of $l_s$ reveal first of all (Fig.~\ref{fig:E0scaling})
that the criterion $\kappa >1$ and SPS are indeed violated
simultaneously near $E=0$ as well as near $E=-2$. What we find
even more important, however, is that the results of the
calculations for different energies, strengths of disorder, and
lengths of the system, all collapse to a single curve, when
presented in the form $\tau (\kappa )$. Thus $\kappa \equiv
l_{loc}/l_s$, after the proper definition of $l_{s}$, remains a
natural scaling variable for the entire spectrum of the system.

This means that separating negative and positive energies in two
bands is physically significant. Of course, functions $\tau
(\kappa )$ for the vicinities of $E=-2$ and $E=0$, do not
coincide. This does not cause problems for our scaling description
as long as these two regions are well separated by the spectral
interval where $\tau =1$.

Additional arguments in favor of our interpretation of the $E=0$
anomaly can be obtained by modifying the initial Anderson model.
For example, one can include next to the nearest neighbors
interaction, and consider a model determined by the equation
\begin{equation}
\psi _{n+1}+\psi _{n-1}+\alpha(\psi _{n+2}+\psi _{n-2})+(\epsilon
_{n}-E)\psi _{n}=0. \label{eq_of_motion1}
\end{equation}
The spectrum of this model, in the absence of disorder, consists
of two branches, only one of which corresponds to propagating
states for $\alpha<1/4$. This branch is invariant with respect to
a similar symmetry $(-1)^n\psi_n(E)=\psi_n(E^{\prime})$. The
stationary point of this transformation $E^{\prime}=E=-2\alpha$ is
a boundary between two adjacent bands. Numerical calculations
confirmed that indeed  there is an anomaly in the dependence of
the scaling parameter $\tau(\kappa)$ at $-2\alpha$ similar to the
one discussed above (see inset in Fig.~\ref{fig:E0scaling}).

\begin{figure}[tbp]
\vspace{-0.3in}
\hspace{-0.4in} % Requires \usepackage{graphicx}
\includegraphics[width=2.9in,angle=-90]{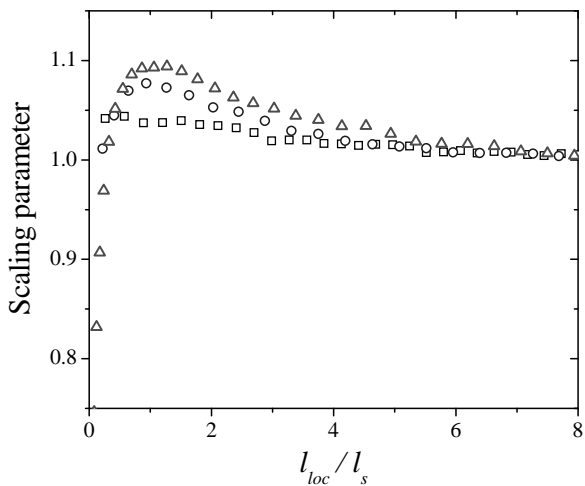}
\caption{ Effect of the periodic potential on the scaling
parameter $\protect\tau$ near $E=0$ for $W = 0.4$ Squares, circles
and triangles correspond to $U_0=0.002$, $0.02$ and $0.2$,
respectively.} \label{fig:Symmetry breaking}
\end{figure}

To justify the physical significance of our treatment of $E=0$ as
a band edge further we added a periodic potential
$U_{n}=U_{0}(-1)^{n}$ to random site energies $\epsilon_n$. This
potential breaks the symmetry of the model, and creates in the
vicinity of $E=0$ a real band gap with a width proportional to
$U_0$.  The idea is to look at the sensitivity of the anomalous
behavior in the vicinity of $E=0$ to the presence of the real band
gap. This gap should decrease together with the magnitude of the
symmetry-breaking potential. Will the features associated with the
gap survive the vanishing of the gap width? The results of the
numerical simulations answering this question are presented in
Fig.~\ref{fig:Symmetry breaking}. One can see that the anomalies
in $\tau (\kappa )$ remain stable with respect to the
symmetry-breaking perturbation. In the limit $U_{0}\rightarrow 0$,
$\tau (\kappa )$ has the same behavior as for small but finite
$U_{0}$. This observation in conjunction with the scaling
arguments presented above gives strong support to our treatment of
$E=0$ as a band edge.

Finally, there is one more question that should be addressed.
Choosing an appropriate periodic potential one can create a band
gap centered not only at $E_{gap}=0$ but anywhere in the spectrum.
Why, then, is SPS violated only near $E=0$? It follows from our
numerical studies that when the periodic potential is weak and
$E_{gap}\ne 0$, the $\tau (\kappa )$-function always approaches
the band edge without any overshoot above its SPS value, $\tau
=1$; it simply decreases inside the gap.  As a result, the normal
SPS-like behavior is restored once the periodic potential
vanishes.  In the Lloyd model, the similar behavior is
characteristic for all band boundaries, $E=\pm 2, 0$. Therefore,
the absence of a real band gap at $E=0$ masks the actual nature of
this energy, and the center-band anomaly in the behavior of the
second moment is not observed.
%One may also ask why the band edge anomaly is absent for the Lloyd
%model even at $E=0$. The answer for this question is similar to
%the answer to the previous one.
%Unlike the Anderson model, $\tau (\kappa )$ in the Lloyd model
%demonstrates no overshoot at the band boundaries. In this respect,
%$E=0$ is not different from other band boundaries, whereas the
%absence of the real gap hides the actual nature of this spectral
%point.

Concluding, we resolved the mystery of the anomalous properties in
the vicinity of $E=0$ in the Anderson model with  diagonal
disorder. We demonstrated that due to an additional symmetry of
the respective Hamiltonian, this spectral point must be considered
as a band boundary between two adjacent bands rather than the
center of a single band. The relevance of the criterion for SPS
suggested in Refs.~\onlinecite{DeychPRL} is thus reestablished.
Our approach allows one to describe statistical properties of
conductance in the vicinity of $E=0$ within the framework of the
scaling approach of Ref.~\onlinecite{DeychPRL2002}.

The authors thank Steve Schwarz for reading and commenting on the
manuscript. The work at Queens College was supported by AFOSR
under Contract No.~F49620-02-1-0305, and partially by PSC-CUNY
grants. BA gratefully acknowledges partial support by the EPSRC
grant GR/R95432.

\end{document}